      [1999/12/01 v1.4c Il Nuovo Cimento]
\documentclass{cimento}

\usepackage{graphicx}  

\usepackage{amssymb}

\title{{Four-top production and $t\overline{t}$ + missing energy events}
{\\ at multi TeV $e^+e^-$ colliders}}
\author{Marco Battaglia\from{ins:1}\from{ins:2} and G\'eraldine ~Servant\from{ins:1}\from{ins:3}}
\instlist{\inst{ins:1} CERN Physics Department, CH-1211 Geneva 23, Switzerland  
\inst{ins:2} University of California, Santa Cruz Inst.\ of Part.\ Phys.\, Santa Cruz, CA 95064, USA
\inst{ins:3} Institut de physique th\'eorique, CEA/Saclay, 91191 Gif-sur-Yvette c\'edex, France}
\PACSes{\PACSit{12.60.cn} \ \ ,  {29.20.db}}

\begin{document}

\maketitle

\begin{abstract}
Four-top production and top pair production in association with missing energy at $e^+e^-$ colliders are sensitive probes of beyond-the-Standard-Model physics. 
We consider Standard Model (SM) extensions containing a new $U(1)^{\prime}$  which couples preferably to the 
most massive states of the SM such as the top quark or Dark Matter but has suppressed couplings to all 
the light states of the SM, as inspired by Randall--Sundrum-like setups or theories of partial fermion 
compositeness.  These simple models are poorly constrained by experimental data but lead to striking new 
signatures at colliders. In this note we consider $Z^\prime$ production in association with a  top quark 
pair in 3 TeV $e^+e^-$ collisions at CLIC, leading to interesting  four-top final states and 
$t\overline{t}+E_{miss}$ events.
\end{abstract}

\section{Theoretical framework}

We introduce a very simple effective theory in which the Standard Model is supplemented by a spontaneously broken $U(1)^\prime$ gauge symmetry, the massive
gauge boson of which acts as a portal to the Standard Model (SM) by coupling to the top quark.  
This type of setup arises naturally in models of  ``partial compositeness" in which the top quark
acquires its large mass (after electro-weak symmetry breaking)  through large 
mixing with composite states in a new strong sector, as in 4d duals to Randall-Sundrum (RS) Models \cite{Randall:1999ee}.
The effective Lagrangian
contains  \cite{Jackson:2009kg},
\begin{eqnarray}
{\cal L}  & = &  {\cal L}_{SM}   - \frac{1}{4} {F}^\prime_{\mu \nu} {F}^{\prime \mu \nu} 
+ M_{{Z}^\prime}^2 {Z}^\prime_{\mu} {Z}^{\prime \mu} 
+ \frac{\chi}{2} \hat{F}^\prime_{\mu \nu} \hat{F}_Y^{\mu \nu} 
+ {g}_t^{Z^\prime} ~ \bar{t} \gamma^\mu P_R {Z}^{\prime}_{\mu}  t 
\nonumber \\ & & ~~~~
+ i \bar{N} \gamma^\mu \left( \partial_{\mu} - i {g}_N^{Z^\prime} P_R {Z}^{\prime \mu}  \right) N
+ M_N \bar{N} N
\label{eq:langrangian}
\end{eqnarray}
where $N$ is a Dirac fermion which is a singlet under the SM gauge interactions but is charged under $U(1)^\prime$ and turns out to play the role of a weakly interacting massive particle (WIMP).
 ${F}^{\prime}_{\mu \nu}$ (${F}^{Y}_{\mu \nu}$) is the usual Abelian field strength for the ${Z}^\prime$ (hyper-charge boson), 
${g}_t^{Z^\prime}$ is the ${Z}^\prime$ coupling to right-handed 
top quarks, and ${g}_N^{Z^\prime}$ is the coupling to right-handed WIMPs.
$M_N$ is the WIMP mass.
One can easily include a coupling to the left-handed top (and bottom).
Our choice to ignore such a coupling fits well with typical RS models
\cite{Randall:1999ee}, balancing the need
for a large top Yukawa interaction with control over corrections to precision electro-weak
observables.
The parameter $\chi$ encapsulates the strength of kinetic
mixing between the $Z^\prime$ and SM hyper-charge bosons.

We have included 
hyper-charge-${Z}^\prime$ kinetic mixing through the term proportional to $\chi$.  
Such a term is consistent with the gauge symmetries, and even if 
absent in the UV,
will be generated in the IR description by loops of 
top quarks\footnote{$\chi$ can be engineered to vanish in the UV,
for example, by embedding $U(1)^\prime$ into a larger
gauge group which breaks down at scales of order $M_{{Z}^\prime}$.}.   
The kinetic mixing parameter $\chi$ generates an effective coupling of SM states to
the ${Z}^\prime$, and through electro-weak symmetry breaking, mass mixing of the
${Z}^\prime$ with the SM $Z$ gauge boson resulting in 
a coupling of $N$ to the SM $Z$ boson. 

 In this approach, most of the Standard Model is fundamental, but with the WIMP, Higgs, 
and right-handed top largely composite.  
The Higgs couples strongly to composite fields, and the amount of admixture in
a given SM fermion determines its mass \cite{Contino:2003ve}.
In this picture, the
$Z^\prime$ is one of the higher resonances, built out of the same preons as the 
WIMP and $t_R$.  RS theories provide a very motivated picture of the UV physics, 
but more generically, in any theory 
containing composite WIMPs 
and composite top quarks 
belonging to a common sector 
one would expect strong couplings between them as a residual of the strong force 
which binds them, and perhaps negligible coupling to the rest
of the Standard Model.

If the $Z^\prime$ mixes with the electro-weak
bosons, this results in strong constraints from precision data.  
We circumvent these constraints by considering a $Z^\prime$ whose mixing with the 
$Z$ is kinetic.  At large $Z^\prime$ masses this is not operationally different
from the mass-mixing case, but it allows us to consider lower mass $Z^\prime$s which
are not ruled out by precision data.

\section{Signals at CLIC}
\label{sec:collider}

\begin{figure}
\begin{center}
\includegraphics[width=0.5\textwidth]{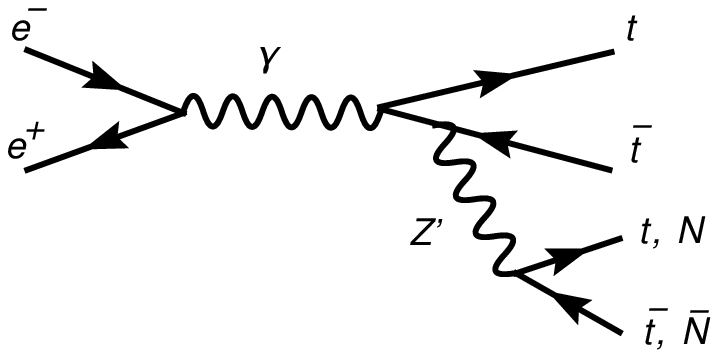}   
\caption{$Z^{\prime}$ mediated production of  $t\overline{t}t\overline{t}$ and 
$ t\overline{t}+E_{miss}$ at CLIC.}
\label{fig:diagram}
\end{center}
\end{figure}
\begin{figure}
\begin{center}
\includegraphics[width=0.5\textwidth]{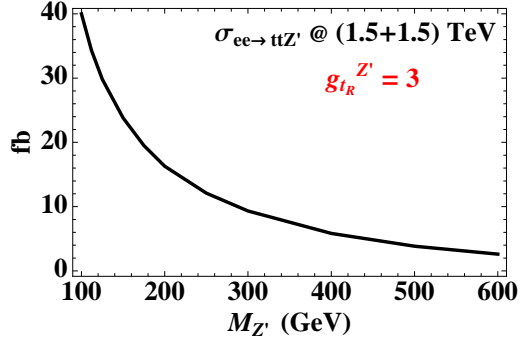}   
\caption{$t\overline{t}Z^{\prime}$ production at CLIC as a function of the $Z^\prime$ mass.}
\label{fig:crosssection}
\end{center}
\end{figure}

Since the coupling of $Z^\prime$ to light SM fermions is suppressed by the 
small kinetic mixing factor, the best probe of the dark sector
is through the top portal.
In particular, the $Z^\prime$ can be produced by being radiated from top quarks, as illustrated 
in Fig.\ref{fig:diagram}. There are very interesting signals at the LHC \cite{Brooijmans:2010tn,gauthier}. 
In this note, we look at the prospects at $e^+e^-$  colliders.
In Fig.~\ref{fig:crosssection}, we show the leading order cross section at CLIC for 
$t\overline{t} Z^\prime$ production, as calculated by CalcHep 2.5.4 \cite{Pukhov:2004ca}.
Depending on the masses and couplings, the $Z^\prime$ will predominantly decay
into $t \overline{t}$, $N \bar{N}$, or into light fermions.
Decays into top quarks lead to four-top events with a very large cross section compared
to the SM four-top rate, which leads to a characteristic same-sign di-lepton
signature \cite{Lillie:2007hd,gauthier} (see also \cite{Contino:2008hi} for studies of a $ttWW$ 
final state at the LHC). 
The right-handed nature of the $Z^\prime$ coupling to top quarks implies top polarisation 
also provides an interesting observable.
When the $Z^\prime$ decays into WIMPs, a $t\overline{t} +$ missing energy final state results,
which is particularly compelling at high energy $e^+e^-$ colliders such as CLIC. 

We present preliminary results of an analysis aimed at characterising these signals in multi-TeV 
$e^+e^-$ collisions. Signal events are generated using CalcHep, assuming 
${g}_t^{Z^\prime}={g}_N^{Z^\prime}=3$, and subsequently hadronised with 
Pythia~6.125~\cite{Sjostrand:2006za}. We assume the beams to be unpolarised. Events are 
processed through full detector simulation using the {\sc Geant-4}-based 
{\sc Mokka}~\cite{MoradeFreitas:2004sq} program and reconstructed with
{\sc Marlin}-based~\cite{Gaede:2006pj} processors assuming a version of the ILD 
detector~\cite{ild}, modified for physics at CLIC. The performances most relevant to the 
analysis of these events are a relative parton energy resolution at 1~TeV 
$\delta_{90} E/E \simeq$~0.08-0.10, charged particle momentum resolution $\delta p_t/p_t^2$ = 
2$\times$10$^{-5}$~GeV$^{-1}$, track extrapolation resolution of 
5~($\mu$m) $\oplus$ $\frac{13 \mathrm{(\mu m GeV^{-1})}}{p_t \mathrm{(GeV)}}$. 
We perform jet clustering forcing the number of jets to the number of top quarks expected 
in the signal final state of interest. Jet reconstruction is performed with the Durham
algorithm~\cite{Catani:1991hj}. In this preliminary study we do not account for machine 
induced backgrounds and do not attempt to perform a detailed top reconstruction through 
the identification of its $W b$ decay, which will need to be evaluated in a more refined 
analysis. We now discuss both signatures in turn.

\subsection{$e^+e^- \to t \bar t + E_{\mathrm{miss}}$}

The Standard Model background for {$e^+e^- \to t \bar t + E_{\mathrm{miss}}$} is dominated by 
 {$e^+e^- \to t \bar t +\nu_e\overline{\nu}_e$}, which, at 3 TeV, has a cross section of 4.1 fb.
 By comparison, the signal cross section for $M_{Z^\prime}=200$  GeV, $M_N=97$ GeV, ${\chi=10^{-3}}$ is 16.5 ~fb (note that this choice of parameters leads to the correct relic density for dark matter \cite{Jackson:2009kg}).   For a higher mass  $M_{Z^{\prime}}= 320$ GeV, the cross section  is 7.6~fb. 
The event selection requires 2 $\le N_{\mathrm{jet}} \le 6$, where $ N_{\mathrm{jet}}$ 
is the number of natural jets obtained with the Durham algorithm for $y_{\mathrm{cut}}$ = 0.0025,  
more than 50 particles with $p_t >$ 0.5~GeV, total reconstructed energy in the 
detector larger than 500~GeV, transverse energy larger than 250~GeV, event thrust 
larger than 0.75 and sphericity lower than 0.30. The missing energy in the event , 
$E_{\mathrm{miss}}$, appears to be the best discriminating variable against the 
SM $t \bar t \nu \bar \nu$ ``irreducible'' background. 
\begin{figure}
\begin{center}
\begin{tabular}{c c}
\includegraphics[width=5.5cm]{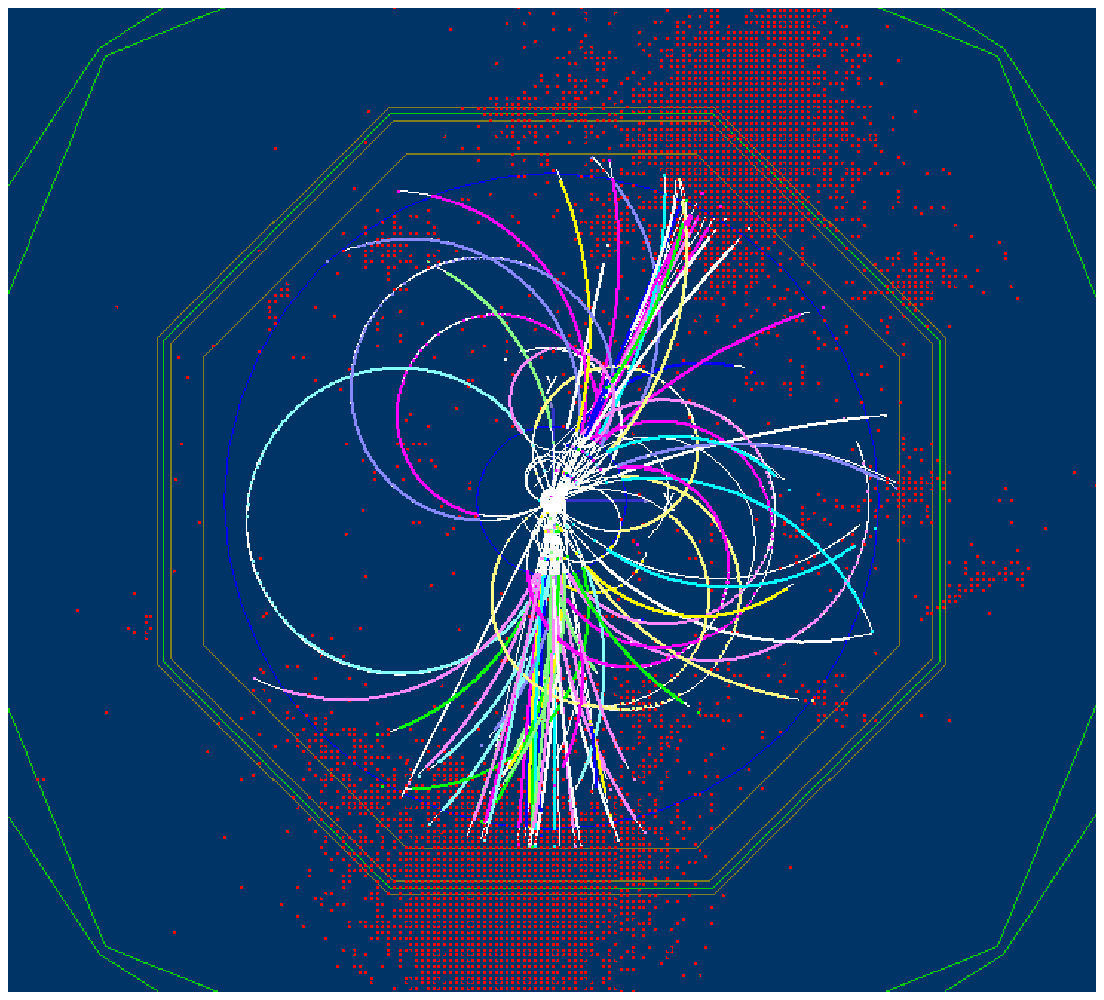} &
\includegraphics[width=7.5cm]{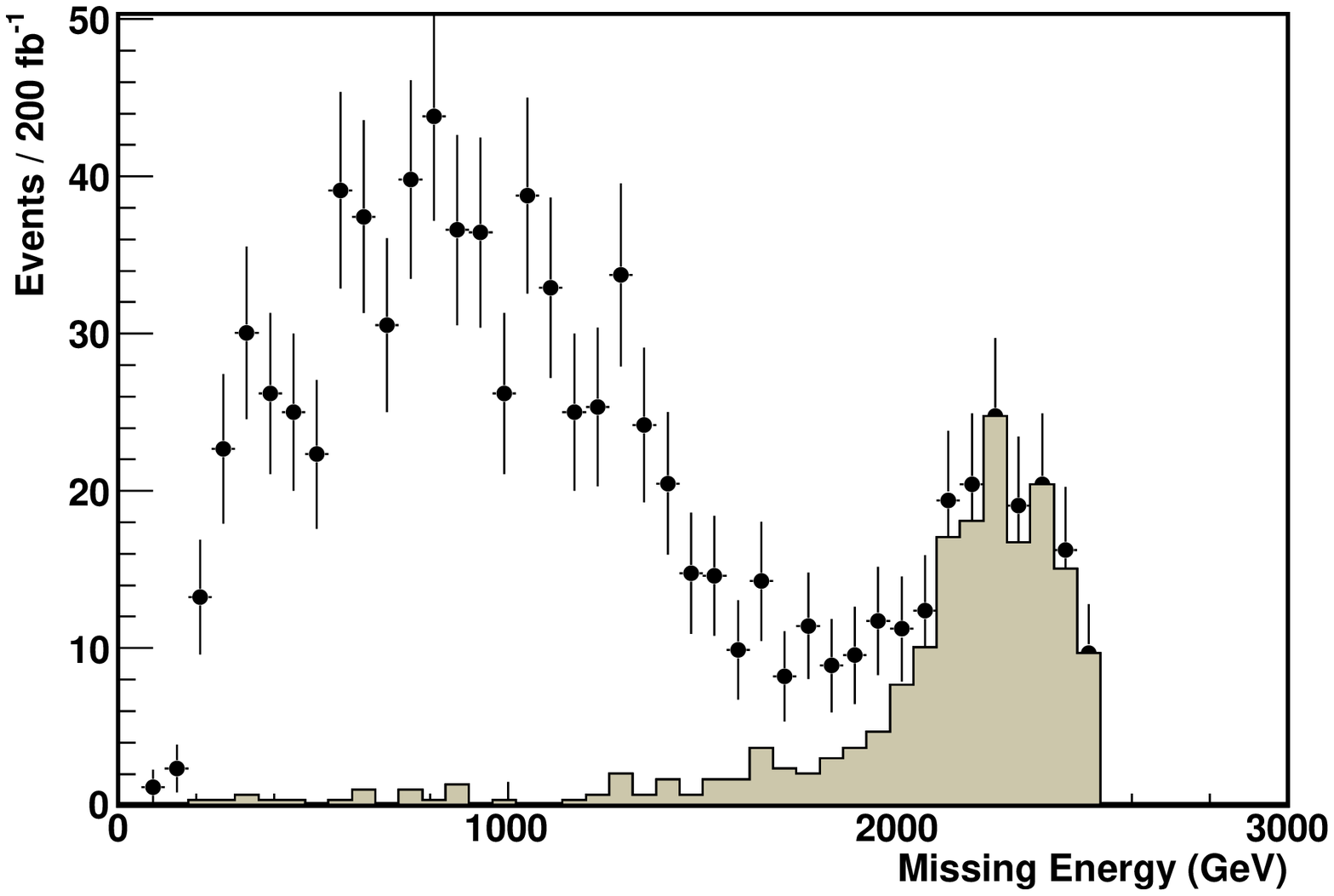} \\
\end{tabular}
\end{center}
\caption{Signal $e^+e^- \to t \bar t Z^\prime \to t \bar t N \bar N$ events at 3~TeV for $M_{Z^{\prime}}= 200$ GeV, $M_N=$ 97 GeV. 
Left: Display of a reconstructed event. Right: Reconstructed $E_{\mathrm{miss}}$ 
distribution. The points with error bars show the reconstructed events for  200~fb$^{-1}$ 
of statistics and the filled histogram the contribution of the SM $t \bar t \nu \bar \nu$ 
background after event pre-selection.}
\label{fig:ttzp1}
\end{figure}
\begin{figure}
\begin{center}
\begin{tabular}{cc}
\includegraphics[width=6.9cm]{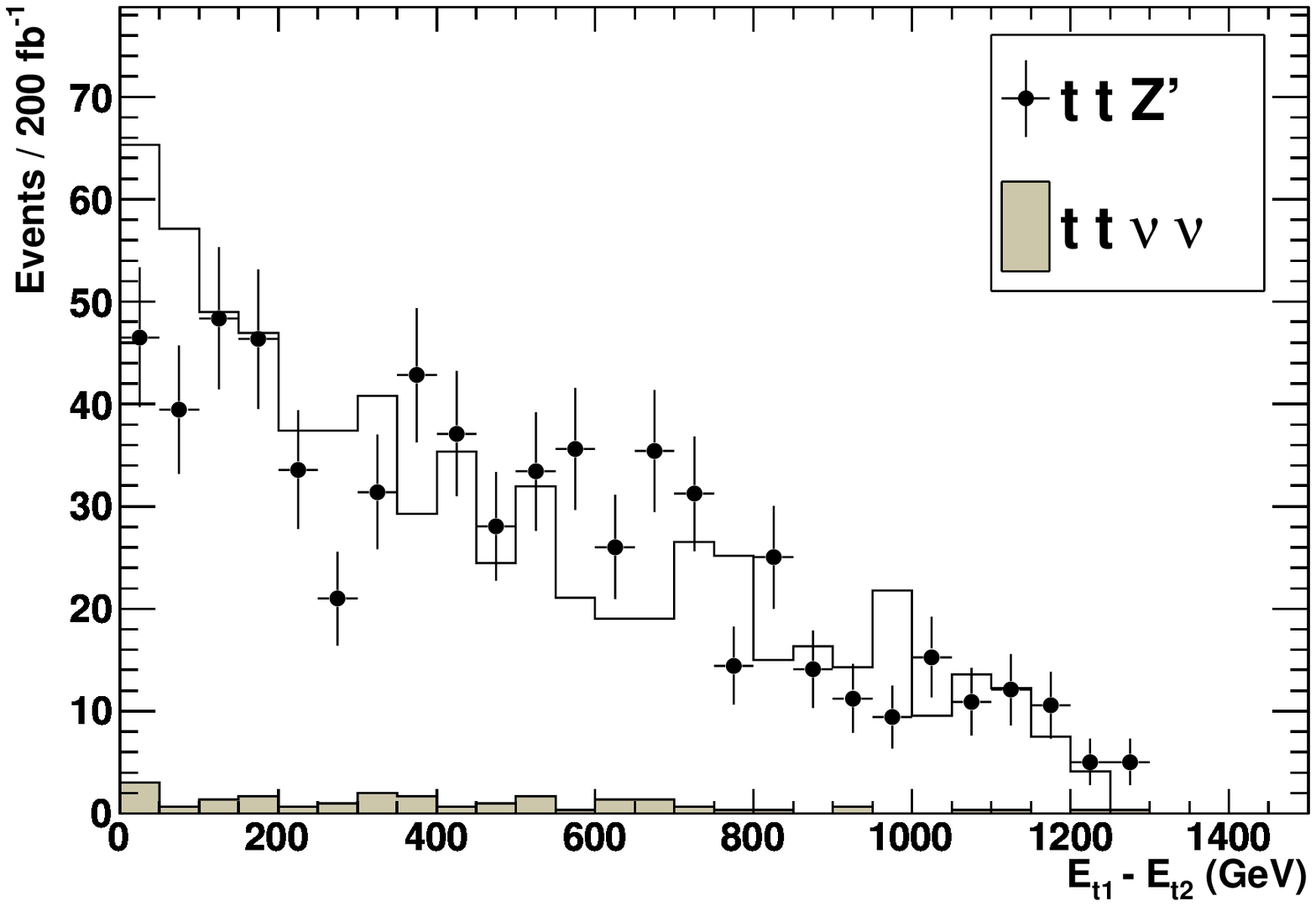} &
\includegraphics[width=6.9cm]{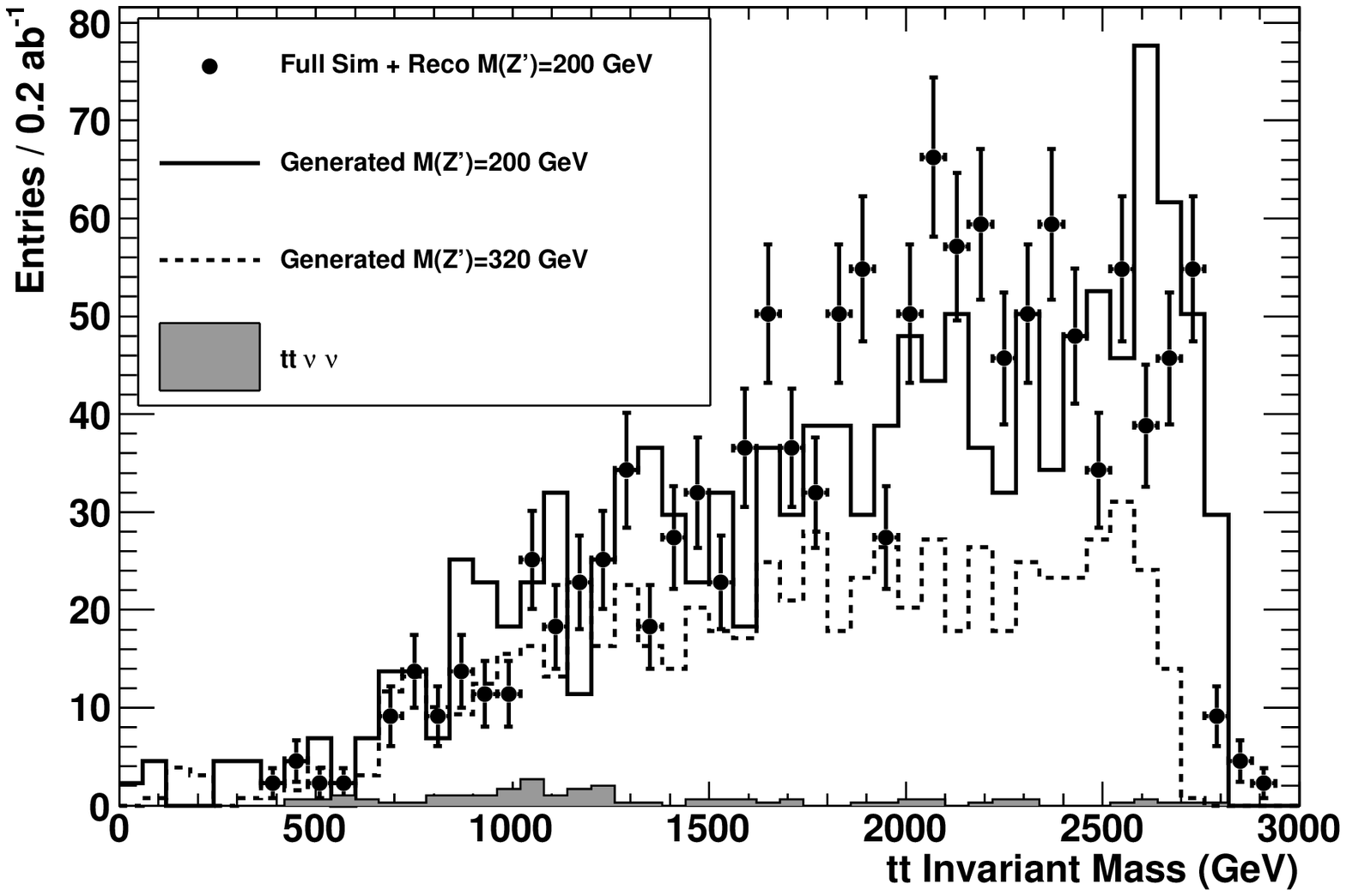} \\
\end{tabular}
\end{center}
\caption{(Left) Difference between the energy of the two top quarks. (Right) Invariant mass of the 
$t \bar t$ pair. The points with error bars show the reconstructed events for  200~fb$^{-1}$ of statistics 
and the filled histogram the contribution of the SM $t \bar t \nu_e \bar \nu_e$ background after event selection. 
The continuous line is the distribution for signal events at generator level. The right panel shows the 
$t \bar t$ invariant mass at generator level for two values of the $Z^\prime$ mass.}
\label{fig:ttzp2}
\end{figure}
Figure~\ref{fig:ttzp1} shows the $E_{\mathrm{miss}}$ distribution for signal and background 
events. We require $E_{\mathrm{miss}} <$ 1900~GeV. Events fulfilling these criteria are 
clustered into two jets. We further require both jets to have $| \cos \theta| <$ 0.90, where $\theta$ 
is the polar angle of the jet axis. Next we compute the invariant mass, $M_{\mathrm{jet}}$ of each 
jet. We require the invariant mass of both jets to be in the range  125~GeV $< M_{\mathrm{jet}} <$ 
225~GeV, around the top quark mass. After this selection, there are 648 signal and 22 SM 
$t \bar t \nu_e \bar \nu_e$ background events, for 200~fb$^{-1}$ of integrated luminosity at 3~TeV. 
Due to the finite parton energy resolution and the beamstrahlung effects, it is not feasible to 
reconstruct the $Z^\prime$ mass from the mass of the system recoiling against the $t \bar t$ pair. 
Instead, we can measure the invariant mass of the $t \bar t$ system. The upper endpoint of this distribution 
is well preserved after reconstruction (see Figure~\ref{fig:ttzp2}) and can be used to estimate the 
$Z^\prime$ mass. 

\subsection{$e^+ e^- \to t \bar t t \bar t$}

\begin{figure}
\begin{center}
\begin{tabular}{c c}
\includegraphics[width=5.5cm,clip=]{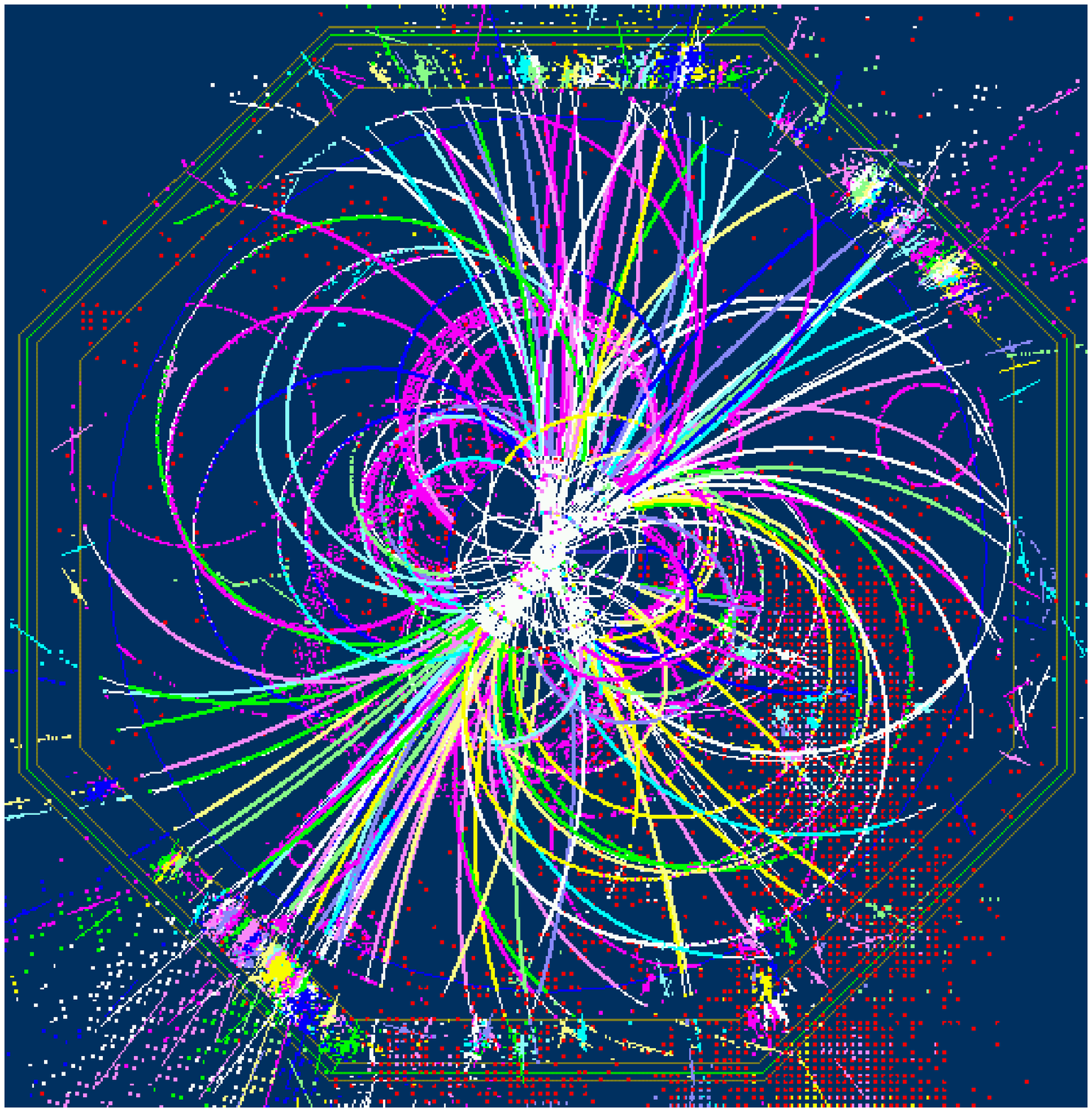} &
\includegraphics[width=7.5cm]{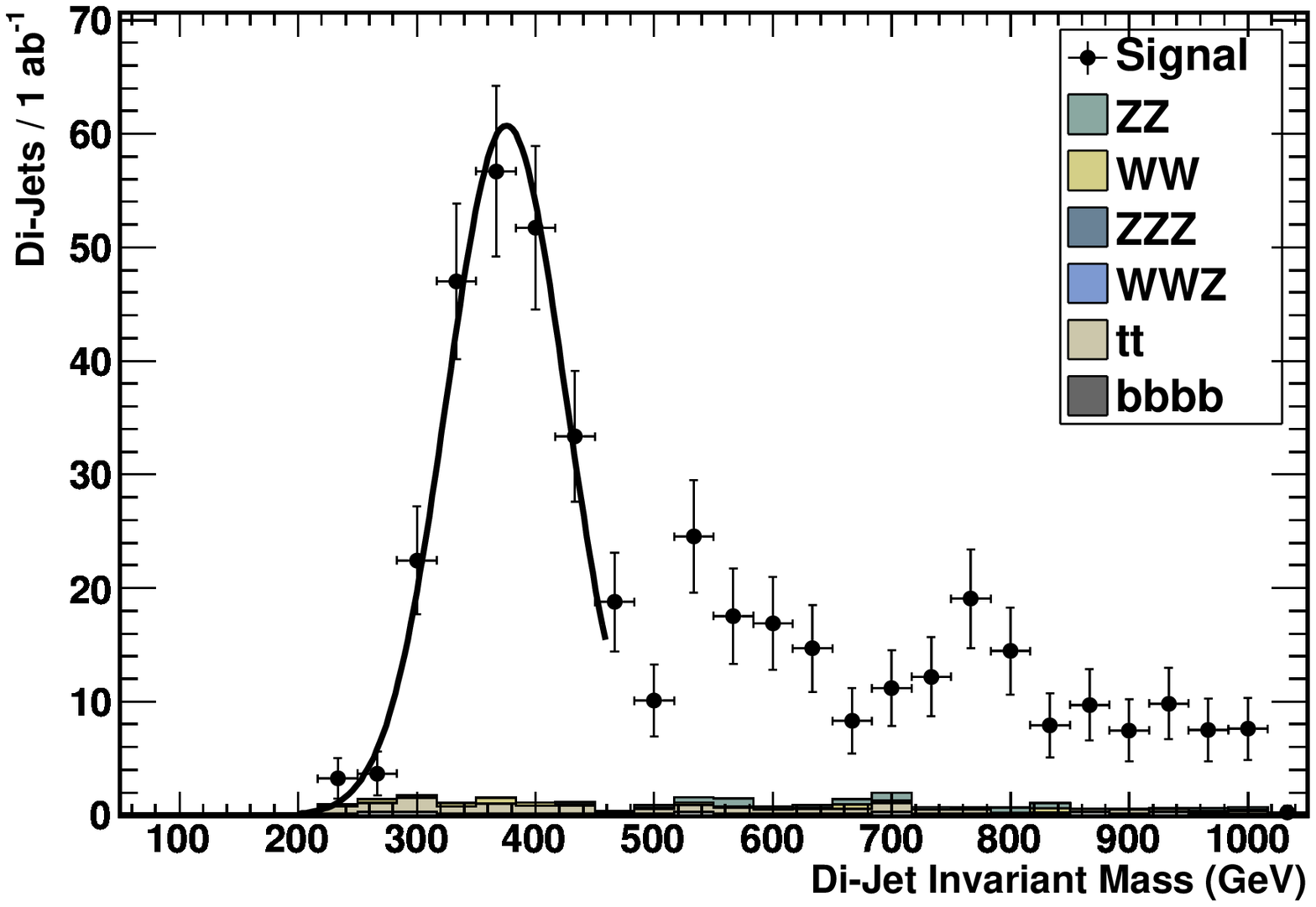} \\
\end{tabular}
\end{center}
\caption{Signal $e^+e^- \to t \bar t Z^\prime \to t \bar t t \bar t$ events at 3~TeV. 
Left: Display of a reconstructed event. Right: Invariant mass of the pair  $t \bar t$ of lowest 
energy in $e^+e^- \to t \bar t t \bar t$ events after kinematic fit with the SM backgrounds 
for 1~ab$^{-1}$ of integrated luminosity and $M_{Z^\prime}$ = 360~GeV.} 
\label{fig:t4}
\end{figure}

The four top quark final state represents a spectacular signature, with minimal background from 
SM processes (see Figure~\ref{fig:t4}). The analysis of the invariant mass of the di-top system 
offers a mean of measuring the $Z^\prime$ mass. The main experimental challenge for this analysis 
is the accurate reconstruction of the twelve parton final state. In this preliminary study, the event 
selection procedure is similar to that adopted for the $ t \bar t + E_{\mathrm{miss}}$ 
channel with the exception that in this case there is no missing energy expected, except through 
leptonic decays in the top decay chain. 

The SM cross section for  $e^+e^- \to t \bar t t \bar t$ at 3 TeV is  0.03 fb while the signal cross section for 
$M_{Z^\prime}=360$ GeV is 4.2~fb. The dominant SM background will actually come from $t\bar t W+ jets$. The cross 
section for $t \bar t Wjj$ is estimated to 0.4 fb while that for $t\bar t WW+N jets$ to 
0.6~fb\footnote{We thank Marcel Vos for providing these estimates.}.
We request events to have more 50 particles with $p_t >$ 0.5~GeV, total reconstructed energy in the 
detector larger than 2.5~TeV, transverse energy larger than 1~TeV, event thrust larger than 
0.96, sphericity lower than 0.50 and more than four jets using Durham clustering.  Events surviving 
these cuts are forced into four jets and a kinematic fit is applied to improve the parton energy 
resolution. We use a port of the {\sc Pufitc} kinematic fit algorithm~\cite{pufitc}.
Figure~\ref{fig:t4} shows the $t \bar t$ invariant mass distribution for 1~ab$^{-1}$ of integrated 
luminosity at 3~TeV. The signal is clearly visible over a negligible SM background. For 
$M_{Z^\prime}$ = 360~GeV there are 240 signal events in the mass peak around the generated 
$Z^\prime$ mass. The $Z^\prime$ mass can be determined with a statistical accuracy of $\simeq$~3.5~GeV.
  
 To conclude, $t\overline{t}+E_{miss}$ and multi-top production are expected signatures for several processes at, and beyond, the TeV frontier. They have the advantage that the irreducible SM background is small so that they can provide clean probes of new physics. They also carry experimental challenges from the reconstruction of a large number of partons, with missing energy and leptons. The prospects for studying these signatures at the LHC are promising \cite{gauthier}.
In this contribution, we started investigating these signatures at $e^+e^-$ colliders using a template model but it is clear that this work can be useful to other models having similar final states.

\acknowledgments
We acknowledge useful discussion with Jean-Jacques Blaising, Marcel Vos and Ambroise 
Espargili\`ere.
We are grateful to the organisers of the LC09 workshop in Perugia for their invitation to an interesting 
and well-organised meeting. The work of G.S is supported by the European Research Council Starting Grant 
Cosmo@LHC.

\end{document}